\documentclass[12pt]{article}
\pdfoutput=1
\usepackage{epsfig,amsfonts,amsthm}
\usepackage{amsmath,amssymb}
\usepackage{mathrsfs}
\usepackage{ae,aecompl}
\usepackage[usenames,dvipsnames]{color} 
\usepackage{slashed}
\usepackage{hyperref}
\hypersetup{
   colorlinks=true,       
    linkcolor=Blue,          
    citecolor=Plum,        
    filecolor=magenta,      
    urlcolor=YellowOrange           
}

\newcommand{\be}{\begin{equation}}
\newcommand{\ee}{\end{equation}}
\newcommand{\bea}{\begin{eqnarray}}
\newcommand{\eea}{\end{eqnarray}}

\providecommand{\Z}{\mathbb{Z}}

\providecommand{\RR}{{\mathbb{R}}}

\providecommand{\bbe}{\mathbf{e}}
\providecommand{\bn}{\mathbf{n}}
\providecommand{\bu}{\mathbf{u}}

\providecommand{\bd}{\mathbf{d}}
\providecommand{\tbd}{\tilde{\mathbf{d}}}
\providecommand{\td}{\tilde{d}}
\providecommand{\bp}{\mathbf{p}}
\providecommand{\bq}{\mathbf{q}}
\providecommand{\tD}{\tilde{D}}
\providecommand{\bss}{\mathbf{s}}
\providecommand{\ts}{\tilde{\mathbf{s}}}

\providecommand{\tphi}{\tilde{\phi}}
\providecommand{\snf}{\mathrm{SNF}}
\providecommand{\yukd}[1][]{\Gamma^{(#1)}}
\providecommand{\yuku}[1][]{\Delta^{(#1)}}
\DeclareMathOperator{\diag}{\mathrm{diag}} 
\providecommand{\eq}[1]{\begin{equation} #1 \end{equation}}
\providecommand{\eqali}[1]{\begin{equation}\begin{aligned} #1
    \end{aligned}\end{equation}}
\providecommand{\bs}[1]{\boldsymbol{#1}}
\usepackage{bbm}
\providecommand{\id}{{\mathbbm{1}}} 
\providecommand{\mss}[1]{\mbox{\scriptsize $#1$}}
\providecommand{\ml}[1]{\mbox{\large $#1$}}
\providecommand{\tp}{{\mss{\mathsf{T}}}}
\providecommand{\ums}[2][1]{\ml{\tfrac{#1}{#2}}} 
\providecommand{\mtrx}[1]{\begin{pmatrix} #1 \end{pmatrix}}
\providecommand{\ZZ}{\mathbb{Z}}
\providecommand{\xlink}[1]
  {\href{http://arxiv.org/abs/#1}{arXiv:#1}}

\def\lsim{\mathrel{\rlap{\lower4pt\hbox{\hskip1pt$\sim$}}
    \raise1pt\hbox{$<$}}}         
\def\gsim{\mathrel{\rlap{\lower4pt\hbox{\hskip1pt$\sim$}}
    \raise1pt\hbox{$>$}}}         

\title{Abelian symmetries of the $N$-Higgs-doublet model with Yukawa interactions}

\author{I.~P.~Ivanov$^{1,2}$, C.~C.~Nishi$^{3}$  
\\
  {\small $^1$ IFPA, Universit\'{e} de Li\`{e}ge, All\'{e}e du 6 Ao\^{u}t 17, b\^{a}timent B5a, 4000 Li\`{e}ge, Belgium}\\
  {\small $^2$ Sobolev Institute of Mathematics, Koptyug avenue 4, 630090, Novosibirsk, Russia}\\
  {\small $^3$ Universidade Federal do ABC - UFABC, 09.210-170, Santo Andr\'e, SP, Brasil}
  }

\begin{document}
\maketitle
\bigskip
\begin{abstract}
We investigate finite abelian groups which can represent symmetries of the $N$-Higgs-doublet models with quarks.
We build a general formalism based on the powerful method of the Smith normal form
and obtain an analytic upper bound on the order of abelian symmetry groups for any $N$.
We investigate in detail the case $N=2$ and rederive known results in a more compact and intuitive fashion.
We also study the NHDM with the maximal finite abelian symmetry for all small $N$ cases up to $N=5$,
and show that in each case all Yukawa textures compatible with such symmetry originate from a unique basic structure.
This work opens the way to a systematic exploration of phenomenology of the NHDM
with a desired symmetry,
and illustrates the power of the Smith normal form technique.
\end{abstract}

\newpage

\tableofcontents

\newpage
\section{Introduction} \label{section-introduction}

\subsection{Two attitudes toward flavor symmetry search}

There is a strong desire to explain the masses and mixing of fermions
by some natural symmetry arguments. 
An ideal outcome of this intense search for a hidden symmetry in the flavor and neutrino sectors
would be to establish a symmetry group whose group-theoretic properties 
would naturally drive all fermion masses and mixing patterns to their values observed in experiment.
In pursuit of this goal, many different groups, frequently finite groups, 
have been proposed and incorporated in various ways, see recent reviews \cite{groups-reviews}.
These groups emerge from extensive scans of finite groups \cite{group-scans},
or from a desire to impose the smallest finite groups with a specific property,
either abelian (such as $\Z_4$ symmetry which drives NNI quark Yukawa textures, \cite{Branco-NNI}) 
or non-abelian (e.g.\ the famous group $A_4$ generating tribimaximal neutrino
mixing, \cite{A4TBM}).

In these approaches one often regards the flavor symmetries as the primary concern
and additional fields as an auxiliary component of the model. 
In particular, one usually does not impose any {\em a priori} restriction on the number or properties
of the scalar fields (Higgs doublets, Higgs singlets, or flavons) 
which are introduced and tuned at will to guarantee the appropriate flavor symmetry breaking.

One can also take another attitude towards incorporating flavor symmetries in
model building. 
Namely, one can first fix the class of models
beyond the Standard Model (bSM) one wants to work with, 
and then systematically study all possible sorts of flavor symmetry groups
which can arise in these models and the phenomenological consequences they lead to. 
On the one hand, from the very start, this approach restricts one's freedom in constructing models.
On the other hand, it can provide the complete classification
of symmetry patterns which are realizable within a certain class of models.
Such theorem-level results are of valuable phenomenological information and can be
used to develop intuition and guide the search for the most efficient symmetry
group.

\subsection{The goals of this work}

In this paper, we take this second attitude in the $N$-Higgs-doublet models (NHDM),
a broad and actively studied class of rather conservative bSM Higgs mechanisms with rich phenomenology.
We assume that there are $N$ copies of scalar doublets $\phi_i$, all having the same electroweak (EW) quantum numbers,
which interact with the gauge bosons and fermions and also self-interact through a certain
EW-invariant renormalizable potential. 

The specific question we study in this paper is which abelian groups can represent symmetries
of the NHDM with quarks for a given number of Higgs doublets $N$, 
and how each allowed symmetry group can be implemented.
The two main goals of this paper are:
\begin{itemize}
\item
to remind the reader of the little known but powerful method linking
the Smith normal form (SNF) of an integer-valued matrix to the abelian (or more accurately, rephasing) 
symmetry group of any bSM model with complex fields;
\item
using it, to find out what are the realizable abelian symmetry groups in NHDM with quarks, 
and what kind of textures in the Higgs-quark Yukawa interactions they lead to.
\end{itemize}
We will recover the known results but in a much more economic way, 
and will provide new results in cases where
the traditional methods become very complicated. 
Although we consider only abelian groups here,
our results also restrict, by virtue of Cauchy's theorem, non-abelian finite symmetry groups realizable in NHDM.
Thus, our analysis has an impact on the entire flavor symmetry search activity
within $N$-Higgs-doublet models.

\subsection{Technical remarks}\label{subsection-technical}

Right away, we would like to make some technical remarks in order to eliminate possible misunderstanding.
The reader who is more interested in the results themselves can skip this subsection.

First, the methods developed here work for {\em rephasing} groups, that is,
subgroups of $U(n_F)$ consisting of rephasing transformation of $n_F$ complex fields present in the problem.
One could of course think of other abelian subgroups of $U(n_F)$; however it is known that any such subgroup
can be bijectively mapped to a rephasing group by a certain basis transformation.
Therefore, our results such as restrictions on orders of the finite groups are basis-independent 
and are valid for abelian groups irrespectively of their embedding in $U(n_F)$. 

If one is forced to work with subgroups of groups other than $U(n)$ or $SU(n)$,
we caution against unjustified uses of this ``abelian $\simeq $ rephasing'' equivalence. 
A particular example where it breaks is $PSU(n) \simeq U(n)/U(1) \simeq SU(n)/\Z_n$. 
Here, one gets extra abelian subgroups
of $PSU(n)$ which cannot be mapped to a rephasing group by a $PSU(n)$ basis change.
This situation was encountered in the classification of horizontal symmetry groups 
of the scalar sectors in the multi-Higgs-doublet models,
\cite{abelianNHDM,finite3HDM}.
There, the overall rephasing group $U(1)$ is factored out because
it is a part of the gauge symmetry and does not represent any novel structural symmetry of the scalar sector.

Second remark concerns the term ``realizable'' groups, which we use following
Refs.\,\cite{abelianNHDM,finite3HDM}.
When imposing a symmetry group $G$, one can end up with a model invariant under a {\em larger} class of transformations,
those which form a group $\tilde G \supset G$. In some cases, it does not lead to significant consequences, in other cases it does.
For example, if one works with several scalars and imposes $G$ which is finite but gets $\tilde G$ which is continuous,
then one runs into trouble with unwanted Goldstone bosons.
It makes it obvious that it is the group $\tilde G$, not $G$, that represents the
true symmetry content of the model.
In specific models, it is also possible that the true symmetry group
$\tilde{G}_S$ of the scalar potential (relevant for the appearance of Goldstone
bosons) differs from the true symmetry of the Yukawa interactions $\tilde{G}_Y$ so
that the true symmetry of the full lagrangian is $\tilde{G}_S\cap \tilde{G}_Y$.

We call a group $G$ {\bf realizable} if it is possible to construct a model symmetric under $G$
and not symmetric under any larger group $\tilde G$.
If, instead, it happens that imposing $G$ on any model {\em unavoidably} makes it
invariant under a larger group $\tilde G$, we call $G$ non-realizable.
Since we deal in this work with abelian groups, we adapt this distinction in the following way: 
an abelian group $G$ is called realizable if there exists a model symmetric under $G$ and not symmetric
under any larger {\em abelian} group $\tilde G \supset G$.

In this work, we deal exclusively with realizable (abelian) groups.
In particular, when we find a complete list of finite realizable abelian groups for a certain $N$, 
it means that trying to impose any other abelian group will unavoidably lead to a model with a larger 
(and usually continuous) abelian symmetry.

\subsection{Classifying NHDMs with discrete abelian symmetries:
current situation}

There exist numerous models which implement various discrete flavor groups \cite{groups-reviews}.
Discrete abelian flavor symmetry groups in models with several Higgs doublets 
are also often used. These symmetries are introduced to achieve a variety of purposes
such as naturally suppressed flavor-changing neutral currents\,\cite{nfc},
stabilization of dark matter candidates together with radiative neutrino
mass generation\,\cite{darkmatter}, specific texture-zeros in the quark mass
matrices\,\cite{grimus:texture}.
However, these particular cases do not elucidate the {\em entire} spectrum of
symmetry opportunities in NHDM.
This challenging task has been tackled only recently.

Within the two-Higgs-doublet model \cite{TDLee,review2HDM}, the simplest version of NHDM, 
the vast majority of works deal with the $\Z_2$ symmetry group, or a product of $\Z_2$ groups.
It is known  \cite{2HDMsymmetries} that within the scalar sector only, nothing else is available 
for a realizable abelian symmetry. Note that this restriction applies only to groups 
of Higgs-family transformations, which act identically on the upper and lower components
of the Higgs doublets. 
If we lift this requirement, more symmetries could be available as discrete subgroups of
continuous groups larger than the horizontal group $SU(2)$, \cite{beyondSU2}.

Inclusion of the fermion sector can extend the list of realizable
horizontal groups.
The full classification of realizable symmetries in 2HDM with quark Yukawa interactions was first done recently in \cite{FS2011}.
It demonstrated that $\Z_3$ is also viable but all higher finite abelian groups are non-realizable.
In addition, it contained a detailed list of Yukawa matrices compatible with $\Z_2$, $\Z_3$, and $U(1)$ groups.
It should be stressed that the approach used in \cite{FS2011} is ``bottom-up'': 
it starts with general Yukawa matrices and generic rephasing transformations,
and then by placing requirements on free parameters, it recognizes allowed patterns and relates them to symmetries.

In the very recent work \cite{Serodio:2013gka}, this approach was generalized to $N > 2$. 
Again, the starting point here is an exhaustive classification of Yukawa textures, which are broken into certain classes.
Then, using a very elaborate combinatorial procedure, \cite{Serodio:2013gka} establishes the minimal groups
one has to impose in order to obtain textures of a given type. 
The results can be useful for phenomenological model building if one knows in advance what kind of quark mass matrices
one wants to exploit. 
However, Ref.~\cite{Serodio:2013gka} gives little information on the full list of possibilities in NHDM with a given $N$,
and in particular, it does not discuss which groups can be imposed for NHDM 
with a given $N$ without producing an accidental $U(1)$ symmetry.

There are several aspects of the ``bottom-up'' approach of \cite{FS2011} and \cite{Serodio:2013gka} which
can leave the reader unsatisfied. Significant part of the work relies on extensive classification of 
numerous textures and on brute-force calculations. 
This approach does not reveal the organizing principle which related textures to symmetry groups,
and it offers no intuitive understanding of what is happening in the quark Yukawa sector of NHDM.
The number of possibilities and the difficulty of calculations quickly grow with the number of Higgs doublets $N$.
It all gives an impression that listing symmetry structures in the NHDM with quarks 
is an extremely difficult enterprise and is impossible to understand intuitively.

The aim of the present paper is to change this situation.
We propose an alternative view of the same problem, which is based on a ``top-down'' approach: 
in contrast to \cite{FS2011} and \cite{Serodio:2013gka},
we first understand the allowed symmetry groups and then implement them. We 
obtain an upper bound of the size of the finite abelian group which can be imposed on the model 
without producing an accidental continuous symmetry,
and outline the general procedure of constructing models realizing a given flavor group.
We also study in detail the NHDMs with the maximal finite flavor symmetry group, for all $N$ up to $N=5$,
and show that in all these cases the quark Yukawa textures are determined by {\em only one basic structure}.
We explicitly construct these Yukawa textures for $N=2$ and $N=3$; 
repeating this construction for larger $N$ should be straightforward.

Our work makes the study of abelian symmetries in NHDM not only systematic but also intuitive.
Restrictions on the allowed finite symmetry groups become understandable, 
the relation between a symmetry group and the possible quark Yukawa textures is straightforward,
and for small $N$, it can be done manually, without relying on computer algebra.
All these clarifications can now guide one's intuition in building bSM models with desired flavor symmetries.
The methods can be extended to neutrinos, or to more complicated scalar sectors.

The structure of the paper is the following. In the next Section we will 
describe the Smith normal form technique in its generality and outline
what extra information can be gained in specific models.
In Section~\ref{section-ABC-method-general} we apply this technique to the generic NHDM with quarks
and obtain restrictions on the realizable groups.
In Section~\ref{section-ABC-method} we discuss the problem of actually constructing 
the Yukawa interactions compatible with a symmetry found at the previous step.
Then, in the following three Sections, we consider in detail
the cases of 2HDM, 3HDM, and NHDM with $N=4$ and $5$.
We close the paper with our conclusions.
Two Appendices contain some auxiliary technical details.

\section{The rephasing symmetry group and the Smith normal form}
\label{section-smith}

\subsection{The method}
The key issue in our strategy is derivation of the rephasing symmetry group
of a given Lagrangian.
This task can be solved for an arbitrary bSM model with any field content. 
The result is expressed in terms of the Smith normal form (SNF) of an integer-valued matrix 
characterizing the interaction patterns of the model. In this Section, we present 
this technique in its full generality.
The relevance of the Smith normal form to this task was first
noticed in \cite{schieren}, where it was applied to classification of discrete remnants of the gauge symmetry 
of GUTs. 
Independently, it was developed in \cite{abelianNHDM}, although without mentioning this name, 
and it was applied to classification of rephasing symmetry groups of the scalar sector of NHDM.
Very recently, it appeared, in certain form, in \cite{varzielas} and was explicitly used in \cite{IL2013}   
in discussions of the geometrical $CP$-violation in multi-doublet models.

Let us start by considering an arbitrary Lagrangian involving $n_F$ complex fields,
all of which can be globally and independently rephased. 
We choose a certain ordering of the fields, labeling them with the index $j = 1,\dots,n_F$.
The rephasing group
of the model is $[U(1)]^{n_F}$, and $j$-th factor here is
the group of rephasing of the $j$-th field by $\exp(i\alpha_j)$,
$\alpha_j \in [0, 2\pi)$. 

Suppose that the Lagrangian contains $k$ different phase-sensitive 
interaction terms. Let us denote by $d_{ij}$ the power of the $j$-th field in
the $i$-th interaction term (if the term contains its conjugate, 
we assign $d_{ij}$ to be minus the power of the conjugate field). 
For example, the scalar interaction term $(\phi_1^\dagger \phi_2)^2$
gives $-2$ for the field $\phi_1$, $2$ for the field $\phi_2$,
and zeros for all other fields.
A typical Yukawa term for quarks, $\bar Q_L \phi d_R$,
yields $1$ for the scalar and the $d_R$ fields, and $-1$ for the $Q_L$ field.

Next, let us consider the $k \times n_F$ integer matrix $D = \{d_{ij}\}$.
If we want the Lagrangian to be invariant under a certain
rephasing defined by angles $\alpha_j$, then these angles must satisfy the following system 
of linear equations:
\be
d_{ij}\alpha_j = 2\pi n_i\,, \quad n_i \in \Z\,.\label{system}
\ee 
If we have a Lagrangian and want to find its rephasing symmetry group,
we need to solve this system.

It turns out that the set of solutions remains invariant under the following 
elementary transformations of the matrix $D$:
\begin{itemize}
\item adding one column (or one row) to another column (or row),
\item flipping the sign of one column or one row,
\item permuting columns or rows.
\end{itemize}
This set of transformation is remarkably powerful because by using them
one can bring any integer-valued $D$ to its \textit{Smith normal form}
$D_{\mathrm{SNF}}$,
which is defined as a matrix whose non-zero entries lie on the main diagonal,
\be
D_{\mathrm{SNF}} = {\rm diag} \left( d_1,\, d_2\, \ldots,\, d_r,\, 0,\, \ldots,\, 0
\right),
\label{smith}
\ee
with positive integers $d_i$ such that each $d_i$ is a divisor of $d_{i+1}$.
When $d_i$ is a nontrivial divisor of $d_{i+1}$, several possibilities should
be distinguished for the same product $\prod_i d_i$ as, e.g., $D_\snf=\diag(1,2,4)$
and $D_\snf=\diag(1,1,8)$ both represent valid SNFs.
In eq.~(\ref{smith}), $r=\mathrm{rank}\ D$.
Note that the matrix itself remains rectangular for $k \not = n_F$;
eq.~(\ref{smith}) represents only its main diagonal, while all non-diagonal
elements are zero.

For any integer-valued $D$, there exists a unique SNF $D_{\mathrm{SNF}}$. It is related to $D$ by
\be
\label{RDC}
D = R D_{\mathrm{SNF}} C\,,
\ee
where the $k \times k$ matrix $R$ encodes all manipulations with rows, while the $n_F \times n_F$ matrix $C$
encodes all manipulations with columns. Clearly, $|\det R| = |\det C| = 1$, because each of them can be written
as a product of matrices describing the above elementary transformations, and each such matrix has determinant $\pm 1$.
This means, in particular, that $R$ and $C$ are invertible.
It also means that the original system (\ref{system}) can be transformed into the system of {\em uncoupled} equations
\be
d_j \tilde \alpha_j = 2 \pi \tilde n_j\,,\label{system-uncoupled}
\ee
for $j = 1,\,\dots,\, r$,
where $\tilde \alpha_j$ are linear combinations
of the initial $\alpha_j$, $\tilde \alpha_j = (C \alpha)_j$,
and the $\tilde n_i$ are also integers, $\tilde n_i = (R^{-1} n)_i$.

Now, since $|\det R| = |\det C| = 1$, the set of solutions of
(\ref{system-uncoupled}) is the same as the set of solutions of (\ref{system}), that
is, we do not lose solutions or bring new ones
by performing elementary manipulations. But the solutions of (\ref{system-uncoupled}) are obvious:
the $j$-th equation has solutions $\tilde \alpha_j = 2\pi \tilde n_j/d_j$, which describe the symmetry group $\Z_{d_j}$
if $d_j \not = 0$ (here, $\Z_1$ denotes the trivial group) or $U(1)$ if $d_j = 0$.
This allows us to immediately establish
the rephasing symmetry of the Lagrangian:
\be
G = \Z_{d_1} \times \Z_{d_2} \times \cdots \times \Z_{d_r}
\times \left[ U(1) \right]^{n_F-r}\,.
\ee
The phases $\alpha_j$ of the generators for each of these groups
are calculated from the corresponding $\tilde\alpha_j$ and the matrix $C$.
For example, the generator of $k$-th finite factor $\Z_{d_k}$, $k = 1,\, \dots,\,r$, 
has phases
\be
\label{RDCphases}
\alpha_j = {2\pi \over d_k} (C^{-1})_{jk}\,.
\ee
where we used the convention (\ref{smith}) for ordering of the diagonal elements.

\subsection{Going beyond the general result}\label{subsection-beyond-general}

Finding the Smith normal form is easily algorithmizable. 
There exist computer-algebra packages \cite{packages}
which do it for any input matrix of integers;
these packages are adequate for practical purposes
and for any reasonable number of fields.
In many cases, even a simpler procedure is sufficient.
Indeed, it is often the case that all $U(1)$ symmetries 
can be easily recognized from physical requirements.
One can get rid of them by striking out certain $n_F-r$ columns.
Then, one picks up $r$ independent rows and columns, 
considers the square $r\times r$ matrix $D'$, which is a submatrix of $D$, and
calculates its determinant. One then has
\be
|\det D'| = d_1 d_2\cdot \dots \cdot d_r\,.
\ee
So, if it happens that $|\det D'|$ is an integer whose prime decomposition
involves only first powers of primes, then one can find the unique 
$d_i$ even without finding the Smith normal form and, therefore,
unambiguously reconstruct the symmetry group.
If $|\det D'|$ contains a prime in a higher power, then this simple
procedure is not sufficient, and one needs to actually compute the Smith normal
form.

The above description also makes it clear what is the minimal number $k$ of 
interaction terms one has to include in order to avoid accidental continuous
symmetries.
Suppose that the structure of interactions automatically conserves $n_G$ global $U(1)$ symmetries.
For example, when considering the scalar sector of NHDM, we have $n_G=1$ because
all possible interaction terms are invariant under the simultaneous global rephasing
of all doublets by the same amount (which in fact can be viewed as the manifestation of the $U(1)_Y$ from the EW 
gauge group). If Yukawa interactions with quarks, leptons and Dirac neutrinos are
included, then one has global $U(1)_Y$, $U(1)_{B}$ and $U(1)_L$ symmetries, so
$n_G=3$.
Therefore, $\mathrm{rank}\ D \le n_F-n_G$.
If $k < n_F-n_G$, then in fact $\mathrm{rank}\ D \le k$, which implies {\em extra} global $U(1)$ symmetries.
So, the necessary condition to avoid this situation is to take at least $n_F-n_G$
interaction terms.

We end this Section with some remarks concerning the application of the SNF technique to specific models.
The general strategy described above does not provide immediate answers for 
specific models. At first glance, the task of classifying rephasing symmetry groups
in each model still remains difficult: with a given number of fields, 
one would need to write down the set of all possible interaction terms, 
which can already be long,
and then check {\em all possible subsets} of this set by calculating SNF for each of them.
Obviously, even for simple models with quarks, this set of all subsets is enormous,
and it cannot be worked out by human.

There is, however, an elegant way out. In many cases, when the interactions
are of a certain very specific type, the rows of the matrix $D$ become
very simple, as they involve very few non-zero entries, 
and in addition these entries are small.
Then, by applying some matrix algebra, one might be able to prove certain
results algebraically, without the need of checking all possible combinations.
For example, one can obtain an upper bound on the value of 
$|\det D|$ (and therefore, on the order of the finite group), or prove that 
all groups of certain classes are realizable.

An example of this powerful technique was presented in \cite{abelianNHDM}.
It led to the exact upper bound on finite rephasing symmetry groups in the scalar sector of NHDM 
for {\em any} number of doublets, $|G| \le 2^{N-1}$,
to the proof that all $\Z_p$ with $1 \le p \le 2^{N-1}$ are realizable,
and to other similar results. In short, it almost gave the final answer to the question
``what are all rephasing symmetry groups realizable in the scalar sector of NHDM for any given $N$?''
without even starting to check combinations of interaction terms.

Our analysis of the abelian symmetries in the NHDM with quarks will be conducted in the same spirit.

\section{Abelian symmetries in NHDM with quarks: realizable groups}\label{section-ABC-method-general}

\subsection{General construction}
We now apply the SNF technique to the NHDM with quarks.
We have $N$ Higgs doublets $\phi_{j_\phi}$, three left-handed quark doublets
$Q_{Lj_L}$, and six right-handed quark singlets $d_{Rj_d}$ and $u_{Rj_u}$, 
with $j_\phi = 1,\dots,N$ and $j_L,\, j_d,\, j_u,\, = 1,2,3$.
In total, they make $n_F=N+9$ complex fields, and each of them can be rephased independently. 
We order the fields in the following way:
\be
(\phi_{j_\phi};\, Q_{Lj_L};\, d_{Rj_d};\, u_{Rj_u})\,.
\ee
The quark-Higgs Yukawa interactions are given by the following Lagrangian:
\be
- {\cal L}_Y = \Gamma^{(j_\phi)}_{j_L j_d}\bar Q_{Lj_L} \phi_{j_\phi} d_{Rj_d} + 
\Delta^{(j_\phi)}_{j_L j_u}\bar Q_{Lj_L} \tilde \phi_{j_\phi} u_{Rj_u} +
h.c.\label{yukawa}
\ee
Therefore, each row $d^{(i)} = (d_{i1},\,\dots,\,d_{in_F})$ of $D$
has one of the following forms 
\eqali{
\label{d^i:N}
d^{(i)}&= (\ \ \bd_\phi;\ -\bd_L;\ \bd_d;\ \bs{0})\,,&&\text{for $d_R$ terms,} \cr
d^{(i)}&= (-\bd_\phi;\ -\bd_L;\ \bs{0};\ \bd_u)\,,&&\text{for $u_R$ terms.}
}
Note that we omit here the {\em h.c.} terms because they just duplicate the rows of $D$ with an overall minus sign.
Each of the vectors $\bd_L,\bd_d,\bd_u$ is equal to one of the three canonical basis vectors 
$\bbe_i$ in $\RR^3$
and each $\bd_\phi$ is one of the $N$ canonical basis vectors of $\RR^N$.

In order to avoid automatically massless quarks, we require that all right-handed
quarks $d_{R}$ and $u_{R}$ participate in at least one Yukawa interaction.
We can choose any six Yukawa terms involving each $d_{Rj_d}$ and
$u_{Rj_u}$ (obviously, this choice is not unique) and use them to construct the first six rows of $D$:
the first three rows are associated with each of $d_R$'s and the next three rows
are associated with each of $u_R$'s.
The rest of $D$ is filled with the remaining $k-6$ terms involving $n_d$ terms with $d_R$'s, 
which we list first, and $n_u$ terms with $u_R$'s; $n_d + n_u = k - 6$.
With all these conventions, we write the matrix $D$ in the following form:
\eq{
\label{D:ABC}
D_{k\times n_F}=\left(\begin{array}{c|c}
  B_{6\times(N+3)} & \id_6 \cr
  \hline
  A_{(k-6)\times (N+3)} & C_{(k-6)\times 6}
\end{array}\right)\,.
}
Applying the elementary transformations, one can bring it to the following form:
\eqali{
\label{D:reduced}
D_{k\times n_F} \sim
\left(\begin{array}{c|c}
  0_{6{\times}(N+3)} & \id_6 \cr
  \hline
  A-CB & 0_{(k-6){\times}6}
\end{array}\right)
\,.
}
The first six entries in $D_{\mathrm{SNF}}$ are obviously $1$'s,
while the remaining $k-6$ entries are the same as in the {\em reduced matrix}
\eq{
\label{tD}
\tD_{k-6}=(A-CB)_{(k-6){\times}(N+3)}\,.
}
The matrices $A$, $B$, and $C$ are naturally broken into two blocks describing $d_R$
and $u_R$ terms:
\eq{
A=
\left(\begin{array}{cc}
  A_d\cr
  \hline
  A_u
\end{array}\right)\,,\quad
B=
\left(\begin{array}{cc}
  B_d\cr
  \hline
  B_u
\end{array}\right)\,,\quad
C=
\left(\!\begin{array}{c|c}
  \bbe_{l_1} & \bs{0} \cr
  \vdots & \vdots \cr
  \bbe_{l_{n_d}} & \bs{0} \cr
  \hline
  \bs{0} & \bbe_{l_1} \cr
  \vdots & \vdots \cr
  \bs{0} & \bbe_{l_{n_u}} \cr
\end{array}\!\right)\quad
\Rightarrow
\quad
\label{CB}
CB=
\left(\begin{array}{cc}
  B_d^{(l_1)}\cr
  \vdots\cr
  B_d^{(l_{n_d})}\cr
  \hline
  B_u^{(l_1)}\cr
  \vdots\cr
  B_u^{(l_{n_u})}
\end{array}\right)\,,
}
where e.g. $B_d^{(l_1)}$ corresponds to the $l_1$-th row of $B_d$.
Finally, we can write the reduced matrix as
\be
\label{D:tilde}
\tD_{k-6} = 
\left(\begin{array}{c}
A_d^{(1)}-B_d^{(l_1)}\cr
\vdots\cr
A_d^{(n_d)}-B_d^{(l_{n_d})}\cr
\hline
A_u^{(1)}-B_u^{(l_1)}\cr
\vdots\cr
A_u^{(n_u)}-B_u^{(l_{n_u})}\cr
\end{array}\right)
=
\left(\begin{array}{c|c}
\bp_1 & \bq_1\\
\bp_2 & \bq_2\\
\vdots & \vdots \\
\bp_{k-6} & \bq_{k-6}
\end{array}\right)\,.
\ee
Here, the $\bp_i$ of length $N$ are Higgs-related vectors whereas $\bq_i$ of length $3$ are related to left doublets.
They have the form
\eqali{
\label{vectorspq}
(\bp_i \, |\, \bq_i)=
A_d^{(i)}-B_d^{(l_i)}&=(\ \ \bd_\phi(A)-\bd_\phi(B) \, | \, -\bd_L(A)+\bd_L(B))\,,\cr
(\bp_{i+n_d}\, | \, \bq_{i+n_d})=
A_u^{(i)}-B_u^{(l_i)}&=(-\bd_\phi(A)+\bd_\phi(B)\, | \,-\bd_L(A)+\bd_L(B))\,,
}
i.e., $\bp_i$ is $\bs{0}$ or $(1,-1,0,\dots,0)$ up to permutation
while $\bq_i$ is $\bs{0}$ or $(1,-1,0)$ up to permutation.

Notice that while the rows of $A^{(i)}_d$ and $A^{(i)}_u$ are generic, 
the rows $B^{(l_i)}_d$ and $B^{(l_i)}_u$ are, generally speaking, not.
Indeed, the matrix $B$ contains only $6$ rows, and all $B^{(l_i)}$ must be selected
from it.
Therefore, if $k > 12$, then some of $B^{(l_i)}$ must repeat.

\subsection{Backbone structure}\label{subsection-backbone}

The above manipulations show that all the information about the rephasing symmetry
group of the model is encoded in the properties of the matrix $\tD_{k-6}$,
Eq.~(\ref{D:tilde}). In this matrix, we are left with the Higgs doublet and
left-handed quark fields, effectively eliminating the right-handed quark
transformation properties. This matrix highlights the constructive role of the Higgs
and left quark spaces and the auxiliary role of the right quark space.

There can be several different matrices $\tD_{k-6}$ related to each other by some
transformations
but leading to the same symmetry group. There are two groups of such transformations:
\begin{enumerate}
\renewcommand{\labelenumi}{(\roman{enumi})}
\item permutations of rows and/or flipping the overall signs of some rows; 
\item permutations of columns within the space of same type fields.
\end{enumerate}
Transformations of rows are unphysical: their permutations 
amount to reordering the list of Yukawa terms,
while the sign flips imply picking up not a term but its conjugate.
Manipulations of columns are potentially relevant for physics,
because other parts of the overall physics Lagrangian do not have to be permutation
invariant.
But their effect is easily recognizable as it is equivalent to relabeling the left doublets fields $Q_{Lj}$
or Higgs doublets $\phi_j$.

In our analysis, we survey the symmetry aspects of the model in
general, and these aspects are blind to the above mentioned transformations. To take
this blindness into account, we will focus on the {\bf backbone structure},
which is defined as the matrix $\tD_{k-6}$ modulo all transformations of type (i)
or (ii).
In particular, when studying the small $N$ examples below, we will classify
different backbone structures rather than different matrices $\tD_{k-6}$.

The issue of how a given backbone structure translates into a specific model will be dealt with
in Section~\ref{section-ABC-method}.

\subsection{$U(1)$ symmetries}

The next step in the analysis is to identify the two 
global $U(1)$ symmetries which are present for all possible quark Yukawa sectors in NHDM.
These are the hypercharge $U(1)_Y$ and the baryon number conservation $U(1)_B$ groups. 

Let us introduce the notation $\bn_p = (1,1,\dots,1) \in \RR^p$ for the vector filled with units
in analogy with the zero vector $\bs{0}_p = (0,0,\dots,0)$.
Then, any possible matrix $D$ satisfies
\eq{
D\bss_Y=\bs{0},\quad D\bss_B=\bs{0}\,.
}
where
\eqali{
\label{s:Y:B}
\bss_Y&=\left(\bn_N;\, {1 \over 3}\bn_3;\, -{2 \over 3}\bn_3;\, {4\over 3}\bn_3\right)^\tp\,,\cr
\bss_B&=\ums{3}(\bs{0}_N;\bn_3;\bn_3;\bn_3)^\tp\,. 
}
The two vectors $\bss_Y$ and $\bss_B$ generate the two-dimensional nullspace
in the space of all possible rephasing angles $\alpha_j$.
Alternatively, one could choose a different pair of basis vectors in the nullspace, for example,
\eq{
\label{s:Y'}
\bss_Y'=\bss_Y-\bss_B=(\bn_N;\,\bs{0}_3;\,-\bn_3;\,\bn_3)^\tp,\quad
\bss_B'=3\bss_B=(\bs{0}_N;\,\bn_3;\,\bn_3;\,\bn_3)^\tp,
}
which have the advantage of containing only integer components.

Conservation of the hypercharge and the baryon number implies that the Smith normal
form of any matrix $D$ contains at least two zeros on the diagonal.
These zeros do not correspond to any {\em flavor symmetry} of the model; they are
present for any collection of the Yukawa quark-Higgs interaction terms.
So, instead of asking what is the symmetry group $G$ of the model, 
we should ask what is the {\em flavor} symmetry group:
\be
G_F = G/(U(1)_Y\times U(1)_B)\,.
\ee

This information can also be extracted from the reduced matrix $\tilde D_{k-6}$, if
we remove one column in the Higgs sector
and one column in the left quark sector, for example, the first and the last columns of $\tilde D_{k-6}$.
This converts \eqref{D:tilde} into
\eq{
\label{def:hat:D}
\hat{D}_{k-6}=
\left(\begin{array}{c|c}
\hat\bp_1 & \hat\bq_1\\
\hat\bp_2 & \hat\bq_2\\
\vdots & \vdots \\
\hat\bp_{k-6} & \hat\bq_{k-6}
\end{array}\right)\,,
}
where $\hat{\bp}_i$ ($\hat\bq_i$) is obtained from $\bp_i$ ($\bq_i$) by eliminating
the first (last) entry.
Indeed, the basis vectors of the nullspace of the matrix $\tilde D_{k-6}$ can be chosen as
$\tilde \bss_Y' = (\bn_N;\,\bs{0}_3)$ and $\tilde \bss_B' = (\bs{0}_N;\, \bn_3)$,
which justifies this operation. 
The SNF of $\hat{D}_{k-6}$ will then give us $G_F$.
It is clear that other choices of columns to be removed are equally acceptable;
this choice will not affect the SNF of the remaining matrix $\hat{D}_{k-6}$.
This is why we defined the backbone structures in terms of $\tilde D_{k-6}$ rather
than $\hat{D}_{k-6}$.

Now, suppose that we require that the model contains no continuous flavor symmetry ($G_F$ is finite).
Then, the SNF of $\hat{D}_{k-6}$ cannot contain zeros on the diagonal.
In particular, it implies that its rank is equal to $N+1$, the number of columns.
From here, we immediately conclude that in order to avoid continuous flavor symmetries,
the model must contain at least 
\be
k_{min} = n_F-2= N+7 
\ee
interaction terms. This condition is necessary but not sufficient: it is possible to devise 
a model with $k \ge N+7$ terms but which still has an additional flavor $U(1)$
symmetry.
Anticipating discussion of Sect.~\ref{subsection-NNI}, we mention here that 
the 2HDM variant suggested in \cite{Branco-NNI} has 10 Yukawa quarks terms
but nevertheless contains an accidental $U(1)$ symmetry.

If we have $k > N+7$ terms and we made sure that $\mathrm{rank}\ \hat{D}_{k-6} = N+1$, 
we can select an appropriately chosen $(N+1)\times(N+1)$ {\em square} submatrix
of the \emph{same} rank, which we now denote by $\hat{D}_{N+1}$, and proceed with
calculating its determinant.

\subsection{An upper bound for $|G_F|$}\label{subsection-bound}

As already discussed in Section~\ref{subsection-beyond-general}, the fact that the rows of $\hat{D}_{k-6}$
have very simple structure allows one to develop the theory further 
and, in particular, to obtain a useful upper bound on the order 
of the realizable finite groups $G_F$ for any given $N$.

We start with the following simple Lemma.
Consider any rectangular matrix $P$ whose rows 
$\bp_i$ are all of the form $(1,\,-1,\,0,\,\dots,\,0)$ up to permutation.
Then, any of its square submatrix $\tilde P$, which is obtained from $P$
by removing the appropriate number of columns and rows,
has determinant $0$ or $\pm 1$. 
Perhaps, the simplest proof is to actually start calculating the Smith normal form
of $\tilde P$ and observe that it can contain only $1$s and $0$s.

Next, let us consider $\hat D_{n}$, a square submatrix of \eqref{def:hat:D} of size
$n\equiv N+1$.
Let us start calculating its determinant via expansion by minors over the last two
columns:
\be
\det \hat D_{n} = \sum_{i < j} (q_{i1} q_{j2} - q_{j1} q_{i2}) \cdot
(-1)^{i+j+1}\cdot \det
P_{(ij)}\,,
\label{dettD}
\ee
where we have defined $q_{i1}\equiv(\bq_i)_1=(\hat{D}_{n})_{i,n-1}$ and 
$q_{i2}\equiv(\bq_i)_2=(\hat{D}_{n})_{i,n}$.
Here $P_{(ij)}$ is the $(i,n-1;\, j,n)$-th minor of $\hat D_{n}$, that is, 
the square $(n-2)\times (n-2)$ matrix
which is constructed from $n-2$ vectors $\hat \bp_{l}$, where $1 \le l \le n$,
by removing the rows $\hat{\bp}_i$ and $\hat{\bp}_j$.
By applying the above Lemma, we know that $\det P_{(ij)}$ is $0$ or $\pm 1$.
Also, by the same Lemma, $q_{i1} q_{j2} - q_{j1} q_{i2}$ can also be only $0$ or $\pm 1$.
Therefore, each term in (\ref{dettD}) can be $0$ or $\pm 1$.
In total, there are $(n-1)n/2$ terms. 
Therefore, $|G_F| = |\det \hat D_{n}|\le (n-1)n/2$.

This upper bound can be improved. 
Indeed, we know that vectors $\hat \bq_{i}$ can be only of
three types:
$(1,0)$, $(0,1)$ and $(1,-1)$ (if some of them happen to be minus these values,
we just flip the sign of the entire corresponding rows). 
To be specific, suppose that we have
$n_1$ vectors $\hat \bq_{i}$ of type $(1,0)$, $n_2$ vectors of type $(0,1)$, 
and $n_3$ vectors of type $(1,-1)$, with $n_1+n_2+n_3=n=N+1$.
This allows us to compute how many terms in (\ref{dettD}) have non-zero $q_{i1} q_{j2} - q_{j1} q_{i2}$:
there are
\be
\label{nonzero:n}
m= n_1n_2 + n_2n_3 + n_3n_1 = {n^2 - n_1^2-n_2^2-n_3^2 \over 2}
\ee
such terms.
The maximal $m$ is attained when $n_1$, $n_2$, $n_3$ stay as close to each other as
possible:
\be
|G_F| \le {n^2 - n_1^2-n_2^2-n_3^2 \over 2} \le {1 \over
3}n^2\,,\label{upperbound2}
\ee
where the second equality is achieved only if $n$ is divisible by 3.
We do not discuss here whether this upper bound is exact. 
In the following Sections we will find the answer {\em a posteriori} for small $N$,
namely, we will show that this upper bound is indeed exact for all
$N\le 5$ giving examples of the groups $G_F$.

For the first few values of $N=n-1$, we get
\be
\label{TableGF}
\begin{array}{r|ccccc}
N & 2 & 3 & 4 & 5 & 6 \\\hline
|G_F| \le  & 3 & 5 & 8 & 12 & 16 \\
|G_\phi| \le & 2 & 4 & 8 & 16 & 32
\end{array}
\ee
In the last row here we show for comparison the largest order of a discrete
rephasing group $G_\phi$ coming only from the NHDM Higgs potential
\cite{abelianNHDM},
and we know that this is the exact upper boundary.
Let us also list the number $(n_1,n_2,n_3)$ of repeated $\hat{\bq}_i$ vectors in
\eqref{upperbound2} when $|G_F|$ is equal to its upper bound:
\be
\label{TableGF:n}
\begin{array}{c|ccccc}
N & 2 & 3 & 4 & 5 \\\hline
|G_F|  & 3 & 5 & 8 & 12  \\
(n_1n_2n_3) & (111) & (211) & (221) & (222)
\end{array}
\ee

One conclusion we can draw is that for $N \ge 5$, the scalar sector
can produce discrete rephasing symmetry groups $G_\phi$ which are larger 
than the Yukawa quark sector can accommodate.
For 2HDM and 3HDM, the situation is reverse:
this upper bound allows the Yukawa quark sector to have a discrete abelian symmetry group
which does not fit the scalar sector.
In this case the scalar sector, considered alone, would be unavoidably symmetric under
an accidental $U(1)$. 
For 4HDM, the maximal symmetry for Yukawas and scalars coincide and is equal to
$\Z_8$. However, it remains to be explicitly checked whether the $\Z_8$ symmetry
of the entire Higgs plus quark Lagrangian actually produces the $\Z_8$ symmetry in
the scalar sector.

\subsection{Alternative way of calculating $\det\hat{D}$ }
\label{subsection-bound:2}

Let us show here another way of calculating $|\det\hat{D}_{N+1}|=|G_F|$, which
will be useful in the forthcoming sections.

Given the rows $\tbd_i$ of $\tilde{D}_n$, we first recognize that $\det\hat{D}_n$
can be written as 
\eq{
\det\hat D_n=\epsilon_{i_0i_1i_2\cdots
i_n i_{n+1}}\td_{1i_1}\td_{2i_2}\cdots\td_{ni_n}\,,
}
where $\td_{aj}=(\tbd_{a})_j$ and $\epsilon_{i_0i_1i_2\cdots i_n i_{n+1}}$
is the totally antisymmetric tensor in $n+2$ dimensions.
Therefore, $\det\hat D_n$ can be written as an antisymmetric linear function (much
like the determinant) of $n$ vectors in $\RR^{n+2}$, which we denote as
\eq{
\label{def:E}
\det\hat D_n=E(\tbd_1,\tbd_2,\cdots,\tbd_n)\,.
}
Antisymmetry of $E$ means that we get a minus sign by exchanging any two vectors,
e.g.,
\eq{
\label{E:antis}
E(\tbd_1,\tbd_2,\cdots,\tbd_n)=-E(\tbd_2,\tbd_1,\cdots,\tbd_n)\,,
}
and linearity in the first entry means that 
\eq{
\label{E:linear}
E(a\tbd_1+b\bu,\tbd_2,\cdots,\tbd_n)=
aE(\tbd_1,\tbd_2,\cdots,\tbd_n)+bE(\bu,\tbd_2,\cdots,\tbd_n)\,,
}
for any real numbers $a,b$; $E$ is analogously linear in any entry.
These two properties lead to the well known result that
$E(\tbd_1,\tbd_2,\cdots,\tbd_n)=0$ if any number of vectors $\tbd_i$ are linearly
dependent.

Now we decompose $\tbd_i$ as
\eq{
\label{d=p+q}
\tbd_i=\bp_i+\bq_i\,,
}
where we slightly change the notation of \eqref{D:tilde} by redefining $\bp_i$ and
$\bq_i$ as vectors in $\RR^{n+2}$ by including appropriate zero entries as
$\bp_i\to(\bp_i;\bs{0})$  and $\bq_i\to (\bs{0};\bq_i)$. We will use this notation
when dealing with the function $E$, 
and what we mean by $\bp_i,\bq_i$ should be clear from the context.

If we insert \eqref{d=p+q} into \eqref{def:E} and use the linearity property
\eqref{E:linear}, we can expand\linebreak $E(\tbd_1,\tbd_2,\cdots,\tbd_n)$ into
$2^n$ terms.
Only $\binom{n}{2}=\ums{2}n(n-1)$ terms containing two $\bq_i$ and
$n-2$ vectors $\bp_i$ have chance to be nonzero,
because at most two vectors $\bq_i$ and at most $n-2$ vectors $\bp_i$
can be linearly independent.
Among these terms, some will vanish because the two $\bq_i$ vectors will be equal,
except for a sign, since we have only six possibilities for $\bq_i$:
\eq{
\label{set:q+}
\{\pm(\bs{0};1,-1,0),\ \pm(\bs{0};0,1,-1),\ \pm(\bs{0};1,0,-1)\}\,.
}
Moreover, these six vectors lie only on three independent directions which we denote
as distinct types.
Therefore, the number of nonzero terms that remain in the expansion will be limited
by how many ways we can pick up two nonparallel vectors from the $n$ vectors
$\bq_i$.
This number is exactly \eqref{nonzero:n}.
The number of nonzero terms can be further reduced if the $n-2$ vectors $\bp_i$ are
linearly dependent or cancellations occur.
The case where $|E(\tbd_1,\tbd_2,\cdots,\tbd_n)|$ is maximal coincides with the
bound given by \eqref{upperbound2} where the number of vectors $\bq_i$ of each type
$(n_1,n_2,n_3)$ are as close as possible.
The distribution of repeated vectors for small $n=N+1$ can be seen in
\eqref{TableGF:n}.

\section{From symmetry groups to Yukawa textures}\label{section-ABC-method}

The methods described in the previous Section allow one, at least in principle,
to construct matrices $\tilde D_{N+1}$ which correspond to a certain abelian flavor symmetry group $G_F$
within the limits mentioned in (\ref{TableGF}). 
The next problem is how to reconstruct the Yukawa quark-Higgs interaction matrices $\Gamma$ and $\Delta$
(i.e. Yukawa textures) starting from a specific matrix $\tilde D_{N+1}$.
A solution to this problem will provide us with a recipe how to actually build an NHDM 
with quarks realizing the specific abelian symmetry group $G_F$.

When solving this problem, one has to address three tasks: 
\begin{enumerate}
\item
establish all backbone structures which correspond to a given $G_F$;
\item
develop the procedure of reconstructing $\Gamma$ and $\Delta$ out of each backbone
structure; 
\item
investigate how unique this reconstruction is.
\end{enumerate}
Formulating these problems in terms of backbone structures rather than $\tilde
D_{N+1}$ highlights the extra transformation freedom which is present in $\tilde
D_{N+1}$ but which is irrelevant for the symmetry properties of the model,
see discussion in Section~\ref{subsection-backbone}.

In the following Sections, we will solve the tasks 1--3 in several specific cases.
Before going into details, let us discuss caveats in a possible straightforward approach
to these tasks.

\subsection{Difficulties of the straightforward
procedure}\label{subsection-difficulties}

When solving the above tasks, we start with a specific flavor group $G_F$ and end up 
with Yukawa textures $\Gamma$'s and $\Delta$'s. 
A straightforward solution of the above tasks would consist in performing the procedure of Sect.~\ref{section-ABC-method-general}
backwards.
Namely, we first choose all matrices $\tilde D$ which correspond to the group $G_F$;
then we find all matrices $A, B, C$ corresponding to each $\tilde D$, 
and then we read the Yukawa textures $\Gamma$'s and $\Delta$'s off the entries of $A, B, C$.

The big problem with this procedure is that it is immensely redundant.
A single backbone structure leads to multiple $\tilde D_{N+1}$ and to even larger number of $\tilde D_{k-6}$.
A single $\tilde D_{k-6} = A - CB$ leads to multiple choices of $A, B, C$. However, 
many differently looking matrices $A, B, C$ (that is, differently looking matrices $D$) lead to the {\em same} model
with different ordering of the terms in the lagrangian.
Indeed, when breaking $D$ into matrices $A, B, C$, we have much freedom to select the six rows which will represent 
the matrix $B$.
Thus, for any given flavor group $G_F$, the overall situation can be schematically represented as
\be
\mbox{few backbone structures} \quad \longleftrightarrow \quad \mbox{very many $A$, $B$, $C$'s} \quad \longleftrightarrow \quad \mbox{few models}\,.
\ee
It is the huge amount of $A$, $B$, $C$'s that makes it difficult to directly map backbone structures  
to the models with their quark Yukawa matrices. Therefore, a more efficient method should avoid 
explicit manipulation with matrices $A$, $B$, $C$.

\subsection{$G_F$ charges}

There might exist a direct and universal method to link the backbone structure with the set of models 
for any group $G_F$, which would bypass matrices $A$, $B$, $C$ altogether. We could not find one. 
However, we found a method which avoids massive redundancy of the above straightforward calculation
and works fine in the small-$N$ cases with large symmetry groups $G_F$. 
This method uses the charges $s_j$ of the fields under the flavor group $G_F$.
Before we turn to these examples,
let us make some preliminary remarks concerning these charges.

Since the group $G_F$ is abelian, the charges are additive. Symmetry under $G_F$
implies that 
the sum of these charges remains unchanged at all interaction vertices.
If $G_F$ is a cyclic group $\Z_p$, then $s_j$ are just integers modulo $p$. If instead it is a direct product
of several cyclic groups, $G_F = \Z_{p_1} \times \dots \times \Z_{p_m}$, then charges $s_j$ are $m$-tuples
of integers modulo $p_i$. To simplify the explanations, we will use notation which alludes to the case of 
a single cyclic group $\Z_p$; modifications for the case of product of cyclic groups will be obvious.

Next, let us denote the vector of charges $\bss$ as
\eq{
\label{gen}
\bss=\mtrx{\bss_\phi\cr\bss_L\cr\hline\bss_d\cr\bss_u}\,.
}
Being a generator of the symmetry, this vector annihilates the matrix $D$:
\be
D \bss = 0 \mod p\,.
\ee
When we switch from $D$ to the basis \eqref{D:reduced}, 
we first manipulate with rows, which conserve $\bss$, 
and only at the last stage we manipulate the columns.
Therefore, in the basis \eqref{D:reduced}, the same generator has the following form:
\eq{
\bss'=\mtrx{\bss_\phi\cr\bss_L\cr\hline
\mtrx{\bss_d\cr\bss_u}+B\mtrx{\bss_\phi\cr\bss_L}}\,.
}
The right quark space components of $\bss'$ must be equal to zero, which allows us to recover
the right quark charges from the Higgses and left quarks' charges.
We also note that the Higgs and left quark charges are the same in the matrix $D$ and in the reduced matrix $\tilde D$,
and are equal to
\eq{
\tilde{\bss}=\mtrx{\bss_\phi\cr\bss_L}\,.
}
These charges can be established by the analysis of $\tilde D$.
The full vector of charges in the original basis can be then represented as
\eq{
\bss=\mtrx{\tilde{\bss}\cr-B\tilde{\bss}}\,.
}
We see that the full set of charges depends only on the backbone structure and
the expression for $B$. The massive ambiguity in definitions of $A$ and $C$ becomes
irrelevant.

\section{Yukawa symmetries in 2HDM}

\subsection{Possible groups $G_F$}

In Ref.\,\cite{FS2011}, the problem of finding all abelian symmetry groups $G_F$
in the 2HDM and establishing all possible textures of $\Gamma$'s and $\Delta$'s 
for each group was solved by the direct explicit check of all possible combinations
of rephasing angles $\alpha_j$.
The following conclusions were reached there: (1) $\Z_2$ symmetry can be present
in the Higgs and in the Yukawa sector, (2) $\Z_3$ symmetry can be present
only in the full Lagrangian; the Higgs potential alone then becomes $U(1)$
symmetric;
(3) trying to impose any $\Z_p$ with $p > 3$ automatically produces a model
whose Lagrangian is $U(1)$ symmetric. In short, the 2HDM with quarks can support
only two discrete abelian groups: $\Z_2$ and $\Z_3$.
In what concerns the Yukawa textures, 34 distinct textures (up to permutations of generations) 
were identified, and only 7 of them correspond to the $\Z_3$ case.

As the first illustration of our technique, we now aim at rederiving (some of) these results
in a more economic and transparent way, focusing on discrete symmetry groups.

We start with the generic minimal matrix $D$ without extra $U(1)$ symmetries; 
it has dimensions $k \times n_F = 9 \times 11$.
The values of $\bd_\phi$ in (\ref{d^i:N}) are $\pm(1,0)$ or $\pm (0,1)$.
The resulting matrix $\tilde D_3$ is a $3\times 5$ matrix
\be
\tilde D_3 = \left(\begin{array}{c|c}
\bp_1 & \bq_1\\
\bp_2 & \bq_2\\
\bp_3 & \bq_3
\end{array}\right)\,,
\ee
where $\bp_i$ is either $(1,\,-1)$ or $(0,\,0)$,
and $\bq_i$ is as usual either $(1,\,-1,\,0)$ up to permutation, or $(0,\,0,\,0)$
(we used here the freedom to flip signs on any row of $\tilde D_3$). 
The $3\times 3$ matrix $\hat{D}_3$ is obtained from here by removing one of the first two columns 
and one of the last three columns.

All realizable symmetry groups come from the SNF of $\hat{D}_3$.
As we already established in Section~\ref{subsection-bound}, for the 2HDM case, $|G_F| \le 3$, 
so the finite flavor abelian symmetry groups can only be $\Z_2$ and $\Z_3$.
The $\Z_2$ 2HDM is well studied in literature \cite{review2HDM,Aoki2009}, 
so we just need to take a closer look at the $\Z_3$-symmetric 2HDM to exhaust all the possibilities.

\subsection{Yukawa textures in $\Z_3$-symmetric 2HDM}
\label{sec:2hdm:texture}

We begin by establishing all distinct backbone structures for the $\Z_3$ 2HDM.

Our analysis in Sections \eqref{subsection-bound} and
\eqref{subsection-bound:2} shows that $n_1=n_2=n_3=1$, which completely fixes the
choice of vectors $\bq_i$, i.e., they should be all nonparallel. The expansion by
minors (\ref{dettD}) contains only three terms, and therefore all of them must be
equal and non-zero. This means that none of $\bp_i$ can be a zero vector.
By flipping signs, all three of them can be brought to the form $\bp_i = (-1,1)$.
The final step then is to fix the signs of $\bq_i$.
This procedure leads to the following matrices $\tilde D_3$ and $\hat{D}_3$: 
\be
\tilde D_3 = \left(\begin{array}{cc|ccc}
-1 & 1 & -1 & 1 & 0\\
-1 & 1 & 0 & -1 & 1\\
-1 & 1 & 1 & 0 & -1
\end{array}\right)\,,\quad
\hat{D}_3 = \left(\begin{array}{c|cc}
1 & -1 & 1 \\
1 & 0 & -1 \\
1 & 1 & 0 
\end{array}\right)\,,
\label{tildeD3-unique}
\ee
up to transformations (i) and (ii) of Sec.\,\ref{subsection-backbone}.
We arrive at the remarkable conclusion that {\em there exists a unique backbone
structure for the $\Z_3$-symmetric 2HDM}.
Transformations (i) are irrelevant and most of transformations (ii) are also
irrelevant since they can be compensated by transformations (i). For example,
cyclic permutations on $Q_{iL}$ can be compensated by cyclic permutations on the
rows leaving $\hat{D}_3$ invariant. Only exchange of $Q_{iL}$ leads to a relative
sign change between the Higgs and $Q_{iL}$ subspace.

The next task is to derive all possible textures in $\Gamma$'s and $\Delta$'s leading to this backbone structure.
As we explained in Section~\ref{subsection-difficulties}, 
a straightforward method of linking the backbone structure with Yukawa textures
runs into difficulties. Therefore, we implement here another strategy which avoids them. Namely, 
we first derive constraints on the $\Z_3$ charges of $(\phi_{j_\phi},\,Q_{Lj_L})$,
then we use $\bss_Y',\bss_B'$ invariance to eliminate the
charges of $\phi_1$ and $Q_{3L}$, then we list all possibilities for the $\Z_3$ charges in the right quark space $\ts$,
and using them, we write all possible interaction terms compatible with $\Z_3$ and
recover the Yukawa textures.

The explicit form of $\tilde D_3$ given in (\ref{tildeD3-unique}) allows us to write 
the generator of the $\Z_3$ symmetry group in the Higgs+left subspace:
\be
\alpha_j = {2\pi \over 3}\left(\!\begin{array}{c}1 \\ \hline -1 \\ 1\end{array}\!\right) \ \mbox{for } \hat D_3\,,
\quad \alpha_j = {2\pi \over 3}\left(\!\begin{array}{c}0 \\ 1 \\ \hline -1 \\ 1\\
0\end{array}\!\right) \ \mbox{for } \tilde D_3\,,
\label{generatorsZ3}
\ee
where we adopt the convention of positive charge for $\phi_2$.
This is equivalent to assigning the $\Z_3$ charges $\tilde s_j$ to Higgses and left doublets,
which are related with the phases by $\alpha_j = 2\pi \tilde s_j/3$:
\eq{
\label{s:Z3}
\tilde{\bss}^{(1)}_{\ZZ_3}=\mtrx{0\cr1\cr\hline -1\cr1\cr0}\,.
}
Note that extra zeros here are recovered unambiguously because
the first and last columns in $\tilde D_3$ are linearly independent.
Obviously, the conservation of $\bss_Y'$ and $\bss_B'$ in \eqref{s:Y'} 
allows us to shift the numbers in \eqref{generatorsZ3} by adding any
multiple of $\bn_2$ in the Higgs sector or $\bn_3$ in the $Q_L$ sector.
The $\Z_3$ charges for the right-handed quarks would be determined if we knew $B$.

If we allow for permutations of columns on the backbone structure
\eqref{tildeD3-unique} [transformation (ii) of Sec.\,\ref{subsection-backbone}],
we can recognize that, apart from the charge \eqref{s:Z3}, there is only one more
possibility for the charge which is
\eq{
\label{s:Z3:2}
\ts^{(2)}_{\Z_3}=\mtrx{0\cr1\cr\hline 1\cr -1\cr 0}\,,
}
once we conveniently set the charge of $\phi_2$ to be positive and ensure the 
first and last zeros from the conservation of $\bss_Y'$ and $\bss_B'$ in
\eqref{s:Y'}.
Note that $\ts^{(2)}_{\Z_3}$ can be turned into $\ts^{(1)}_{\Z_3}$ by 
the overall sign flip followed by the exchange of the two Higgs doublets 
and addition of $\bn_2$.

Considering the possible charges \eqref{s:Z3} and \eqref{s:Z3:2}, we note that 
the $\Z_3$ charges of the \emph{three left doublets} are all \emph{distinct} and
\emph{exhaust all the charges}.
This fact places strong restrictions on the possible form of $\Gamma$'s and
$\Delta$'s.
Consider, for example, the possible structures of $\Gamma^{(1)}$. An element $\Gamma^{(1)}_{ab}$
can be non-zero if the charges in this term sum up to zero (modulo 3):
\be
s_{\phi_1} - s_{L_a} + s_{d_b} = 0 \mod 3 \quad \Rightarrow \quad
s_{L_a} = s_{\phi_1} + s_{d_b} \mod 3\,.
\label{Z3charges-sum-up}
\ee
Whatever the charges of $d_R$ and Higgses are, for every $b$ 
there exists one and only one $a$ satisfying (\ref{Z3charges-sum-up}).
We arrive therefore at the following result: $\yukd[1]$ can
have only three non-zero elements, exactly one in each column. 
Clearly, the same argument applies to $\yukd[2]$, and to $\Delta$'s.
Therefore, we conclude that {\em any $\Z_3$-symmetric 2HDM with quarks can
have at most 12  (and at least 9) Yukawa interaction terms, one term per each
combination of a Higgs and a right-handed quark, 6 for $d_R$ and 6 for $u_R$}.

Next, we want to list all combinations of $\Gamma$'s and $\Delta$'s 
with the {\em maximal} number of Yukawa terms (i.e.\ 12) compatible with the $\Z_3$
symmetry in 2HDM.
We write the matrix $D_{12\times 11}$, corresponding to all the 12
Yukawa terms allowed by the $\Z_3$ symmetry, as
\be
D_{12 \times 11} = \left(\begin{array}{cc|c|c|c}
(1 & 0)_3 &A_1 & \id_3 & 0_3 \\  
(-1 & 0)_3 &B_1 & 0_3 & \id_3 \\  
(0 & 1)_3 &A_2 & \id_3 & 0_3 \\ 
(0 & -1)_3 &B_2 & 0_3 & \id_3 
\end{array}\right)\,,
\ee
where the notation $(1,0)_3$ denotes the collection of three equal rows $(1,0)$.
The $3\times 3$ matrices $A_i,B_i$ are composed of zeros and of a single $-1$ in each row.
They are directly related to the Yukawa coefficients as
\be
\label{AB:yukawa}
\yukd[j_\phi] \sim (A_{j_\phi})^\tp \,,\quad
\yuku[j_\phi] \sim (B_{j_\phi})^\tp \,;
\ee
the symbol $\sim$ means that the nonzero entries are located at the same place.
These matrices are not to be confused with $A$ and $B$ introduced in
Section~\ref{section-ABC-method};
those matrices can be immediately extracted from $D_{12 \times 11}$.
The $6\times 5$ reduced matrix \eqref{tD} becomes
\eq{
\label{tD:AB}
\tD_{6}=\left(\begin{array}{cc|c}
	  (-1&1)_3& A_2-A_1\cr\hline 
	  (-1&1)_3& B_1-B_2
    \end{array}
    \right)\,,
}
where we conveniently flipped the sign of the last three rows.

As generators of the $\Z_3$ symmetry, the multiplication of \eqref{tD:AB} by the
generators in \eqref{s:Z3} or \eqref{s:Z3:2} should result in a column vector
containing only 0 or $\pm3$ (in principle, multiples of 3 are allowed but we can see
there is not enough 1's to generate larger numbers). This result should hold
irrespective of the charges of $d_R,u_R$.
Therefore, for the charges $\ts^{(1)}$, we have only three possibilities for the
rows of $A_2-A_1$ and $B_1-B_2$:
\eq{
\label{A2-A1}
(A_2-A_1)_i\sim (B_1-B_2)_i\in \{(-1,1,0), (1,0,-1),(0,-1,1)\}\,.
}
For the charges $\ts^{(2)}$, we have only to multiply the possibilities by $(-1)$.
We can see that in both cases $(0,0,0)$ is not allowed, so that the rows of $A_2$
and $A_1$ ($B_2$ and $B_1$) can not have the nonzero entry at the same place.
Moreover, for $\ts^{(1)}$, we know that the entries $1$ in \eqref{A2-A1} correspond
to $A_1$ ($B_2$) whereas the entries $(-1)$ correspond to $A_2$ ($B_1$).
Therefore, the nonzero entry of $A_2$ ($B_1$) is always cyclically one entry on the
left of the nonzero entry of $A_1$ ($B_2$).
For the charges $\ts^{(2)}$ we have the opposite relation 
(or, alternatively, we just swap the Higgs doublet index $1 \leftrightarrow 2$).

We then arrive at the general recipe to construct all Yukawa textures allowed by
$\Z_3$ symmetry:
\begin{enumerate}
\item take $\yukd[1]$ and, independently, $\yuku[2]$ such that each of their columns
has exactly one nonzero entry;
\item the texture of $\yukd[2]$ ($\yuku[1]$) is determined from the texture
of $\yukd[1]$ ($\yuku[2]$) by moving all rows cyclically one entry up (for
$\ts^{(1)}$) or down (for $\ts^{(2)}$).
\end{enumerate}
To avoid an automatic massless quark, the textures in step 1 must have entries
in at least two rows. If we count options up to permutations in the right quark
space (that is, disregarding possible reordering of $d_{Rj}$ and $u_{Rj}$), then we
get seven possibilities:
\eq{
\mtrx{*&0&0\cr0&*&0\cr0&0&*}\,,\quad
\begin{array}{l}
\mtrx{*&*&0\cr0&0&*\cr0&0&0}\,,\quad 
\mtrx{0&0&0\cr*&*&0\cr0&0&*}\,,\quad 
\mtrx{0&0&*\cr0&0&0\cr*&*&0}\,,\\[8mm] 
\mtrx{*&*&0\cr0&0&0\cr0&0&*}\,,\quad 
\mtrx{0&0&*\cr*&*&0\cr0&0&0}\,,\quad 
\mtrx{0&0&0\cr0&0&*\cr*&*&0}\,.
\end{array}
}
Up to permutation in the right quark space, 
they correspond to the seven structures $\Gamma^{(1)}$ listed in Eqs.~(89)--(95) of
Ref.~\cite{FS2011}.

To give a concrete example of the recipe, take for $\yukd[1],\yuku[2]$,
\eq{
\yukd[1]=\mtrx{0&0&*\cr*&0&0\cr0&*&0}\,,~~
\yuku[2]=\mtrx{*&*&0\cr0&0&*\cr0&0&0}\,.
}
The allowed textures for $\yukd[2],\yuku[1]$ are either
\eq{
\label{G2D1:1}
\yukd[2]=\mtrx{*&0&0\cr0&*&0\cr0&0&*}\,,~~
\yuku[1]=\mtrx{0&0&*\cr0&0&0\cr*&*&0}\,,
}
or
\eq{
\label{G2D1:2}
\yukd[2]=\mtrx{0&*&0\cr0&0&*\cr*&0&0}\,,~~
\yuku[1]=\mtrx{0&0&0\cr*&*&0\cr0&0&*}\,.
}
For completeness, we can check that the $\Z_3$ charges for the whole theory with
texture \eqref{G2D1:1} or \eqref{G2D1:2} are given respectively by
\eqali{
\bss&=(0,1;-1,1,0;1,0,-1;0,0,-1)^\tp,~~\cr
\bss&=(0,1;1,-1,0;-1,0,1;-1,-1,0)^\tp .
}

The analysis above essentially recovers the results of Ref.\,\cite{FS2011} on $\Z_3$
symmetric textures in a simpler way.

\subsection{$\Z_4$-2HDM with the next-neighbour interaction texture}\label{subsection-NNI}

Ref.~\cite{Branco-NNI} considered a model for quark mass generation 
with the next-neighbour interaction (NNI) structure
\eq{
\label{NNI}
M_d\sim M_u \sim\mtrx{0&*&0\cr*&0&*\cr0&*&*}\,.
}
It was shown in \cite{Branco-NNI} that if one wants to produce such a texture
within multi-doublet models purely from symmetry considerations,
then the minimal realization is the 2HDM with the $\Z_4$ symmetry group.
This model was further studied and generalized for more than two doublets 
in a number of later papers.

In many of these works, the original NNI model was referred to as 
the ``$\Z_4$ model'', see Ref.~\cite{Aranda-Z4} for a typical example.
Unfortunately, this is a misnomer. Results of \cite{FS2011}, confirmed also in this work,
show that there can be no 2HDM with a relevant $\Z_4$-symmetry in the Higgs and quark sector.
When such a symmetry is imposed, it unavoidably leads to a continuous $U(1)$ global
symmetry\footnote{Ref.~\cite{Branco-NNI} rightfully mentions that an accidental $U(1)$
symmetry appears in the scalar potential but it does not stress that the Yukawa
sector is also accidentally $U(1)$-symmetric.}.
$\Z_4$ then is nothing but one subgroup of this $U(1)$; other subgroups $\Z_2$, $\Z_3$, $\Z_5$, etc.
can be easily identified. In our language, the group $\Z_4$ is not realizable, see discussion in Section~\ref{subsection-technical}.

Although one {\em starts} by imposing $\Z_4$, one {\em actually}
obtains a $U(1)$ flavor symmetry.
It is true that $\Z_4$ can be  referred to as the smallest flavor
symmetry group one needs to impose in order to guarantee an extra $U(1)$ symmetry.
But once the model is built, one needs to analyze it as a $U(1)$-symmetric model.
In particular, focusing on $\Z_4$ charges and transformations within this model is misleading.

\section{$\Z_5$-symmetric 3HDM: textures in $\Gamma$'s and $\Delta$'s}

In this section we focus on the maximal abelian symmetry which is possible for the
Yukawa sector of the 3-Higgs-doublet-models. The nonmaximal groups are all realizable
in this case, see Appendix \ref{ap:realizable}, but they will not be treated here.

\subsection{Backbone structure}
\label{sec:3hdm:Z5}

Let us apply the method of Sec.\,\ref{sec:2hdm:texture} of deducing textures
$\Gamma$'s and $\Delta$'s to the $\Z_5$-symmetric 3HDM.
We start determining the backbone structures that lead to the $\Z_5$ symmetry.

The backbone structure comes from the reduced $6 \times 4$ matrix $\tilde D_4$
\be
\label{tD:3hdm:5}
\tilde D_4 = \left(\begin{array}{c|c}
\bp_1 & \bq_1\\
\bp_2 & \bq_2\\
\bp_3 & \bq_3\\
\bp_4 & \bq_4
\end{array}\right)\,,
\ee
where $\bp_i$ and $\bq_i$ are $\bs{0}$ or $(1,-1,0)$ up to permutation.
We want to construct $\hat D_4$ such that $|\det \hat D_4| = 5$.

Let us analyze the possible structures for \eqref{tD:3hdm:5}.
According to \eqref{TableGF:n} and Sec.\,\ref{subsection-bound:2}, $|\det\hat
D_4|=|E(\tbd_1,\tbd_2,\tbd_3,\tbd_4)|$ can be maximal only when two out of the
four $\bq_i$ vectors are parallel.
By reordering and adjusting the overall sign of $\tbd_i$, we can always suppose
$\bq_4=\bq_1$.
The expansion of $\det\hat D_4$ in this case yields
\eq{
\label{D4:exp}
E(\tbd_1,\tbd_2,\tbd_3,\tbd_4)=
(\slashed{1}\slashed{2}34)+(\slashed{1}2\slashed{3}4)+(1\slashed{2}\slashed{3}4)
+(1\slashed{2}3\slashed{1})+(12\slashed{3}\slashed{1})\,,
}
where we use the shorthand notation
$(\slashed{1}\slashed{2}34)\equiv E(\bq_1,\bq_2,\bp_3,\bp_4)$: the slashed numbers
refer to $\bq$-vectors whereas unslashed vectors are $\bp$-vectors.
Note that the term $(\slashed{1}23\slashed{1})$ vanishes automatically.
To get $|\det\hat D_4|=5$ we need all the terms in \eqref{D4:exp} to be nonzero 
and equal since each term can be only $0$ or $\pm1$ due to Lemma in
Sec.\,\ref{subsection-bound}.

Now we turn to the vectors $\bp$.
For $n=4$ ($N=3$) the vectors $\bp$ have a structure analogous to the $\bq$
vectors and should be one of the six vectors
\eq{
\label{n=4:p:set}
\bp\in\{\pm(1,-1,0;\bs{0}),\pm(0,1,-1;\bs{0}),\pm(-1,0,1;\bs{0})\}\,.
}
This means that at least two $\bp$-vectors will be parallel to each other.
Since all five terms in \eqref{D4:exp} must be nonzero, we conclude that 
neither $\bp_1$ nor $\bp_4$ can be parallel to any other vector.
Thus, the only choice is to choose $\bp_2$ an $\bp_3$ parallel, i.e.,
$\bp_3=\pm\bp_2$. 
We will decide the sign by avoiding cancellations in \eqref{D4:exp}.

At this point we can make explicit the fact that only two $\bq$-vectors in
\eqref{set:q+} are linearly independent.
By adjusting the signs, we can choose the distinct vectors $\bq_1,\bq_2,\bq_3$ to
obey
\eq{
\label{q3=q1+q2}
\bq_3=\bq_1+\bq_2\,.
}
Now the last two terms of \eqref{D4:exp} can be simplified to $\pm(1\slashed{2}2\slashed{1})+(12\slashed{2}\slashed{1})$.
They do not cancel only if
\eq{
\label{n=4:3=-2}
\bp_3=-\bp_2\,.
}
Let us expand \eqref{D4:exp} again as
\eq{
\label{D4:exp:2}
E(\tbd_1,\tbd_2,\tbd_3,\tbd_4)=
-2(\slashed{1}\slashed{2}24)-(\slashed{1}\slashed{2}14)
-2(\slashed{1}\slashed{2}12)\,.
}
These three terms only add up if three distinct vectors $\bp_1$, $\bp_2$, $\bp_4$
are chosen to obey
\eq{
\label{n=3:p4}
\bp_4=\bp_2-\bp_1\,.
}
Notice that $\bp_1,\bp_2$ cannot be selected arbitrarily from \eqref{n=4:p:set};
they must be such that \eqref{n=3:p4} is also from the list \eqref{n=4:p:set}.
We then find that $E(\tbd_1,\tbd_2,\tbd_3,\tbd_4)=-5(\slashed{1}\slashed{2}12)$,
and what remains is to make sure that $(\slashed{1}\slashed{2}12) = \pm 1$.
This can be easily done; for example, the choice
\eq{
\label{choice:q12}
\bq_1=(\bs{0};1,0,-1)\,,~~\bq_2=(\bs{0};-1,1,0)\,,
}
and
\eq{
\bp_1=(0,1,-1;\bs{0})\,,~~\bp_2=(1,0,-1;\bs{0})\,,
}
leads to $(\slashed{1}\slashed{2}12)=-1$.
Any other choice can be recovered by relabeling $Q_{iL}$ and $\phi_i$
appropriately and such a process maintains $(\slashed{1}\slashed{2}12)=\pm 1$.
Then $\tilde D_4$ can be considered to have the standard form
\eq{
\tilde D_4=
  \left(\begin{array}{ccc|ccc}
  0 & 1 & -1 & 1 & 0 & -1\\
  1 & 0 & -1 & -1 & 1 & 0 \\
  -1 & 0 & 1 & 0 & 1 & -1 \\
  1 & -1 & 0 & 1 & 0 & -1 
  \end{array}\right)\,.
\label{backbone-Z5:std}
}

To answer the uniqueness question, we recall that 
the only freedom in constructing $\tilde D_4$ left after fixing
the pairs of parallel vectors is the choice of nonparallel pairs $(\bq_1,\bq_2)$
and $(\bp_1,\bp_2)$, constrained by \eqref{q3=q1+q2} and \eqref{n=3:p4}.
Suppose that we choose another $(\bq_1,\bq_2)$ pair. 
Then by permutation within the left quark space it can be brought back
to (\ref{choice:q12}). The same holds for  $(\bp_1,\bp_2)$.
Therefore, in terms of backbone structures,
we obtain {\em only one backbone structure for the $\Z_5$ symmetric 3HDM}. 
The labeling chosen for (\ref{backbone-Z5:std}) simply maximizes the number of zero entries
within the central square matrix $\hat D_4$.

\subsection{Yukawa textures}

Let us now deal with the possible Yukawa textures.
For that purpose, we first extract the $\Z_5$ charges of the $(\phi,\,Q_L)$ fields.
As usual, we choose the central square block in $\tilde D_4$ to represent $\hat
D_4$, find its Smith normal norm and obtain diag$(1,\,1,\,1,\,5)$.
Then, we extract the phases $\alpha_j$ of the $\Z_5$ generator using (\ref{RDCphases}), 
remove the prefactor $2\pi/5$, and supplement them by zeros in the beginning and in
the end, similarly to (\ref{generatorsZ3}).
The result is 
\eq{
\label{Z5charges}
\ts_{\Z_5}=(0,\,2,\,-1;\,2,\,1,\,0)^\tp\,,
}
where calculations are made modulo 5.
The choice $\ts=(0,\,-1,\,-2;\,-1,\,2,\,0)^\tp$ as well as the charges multiplied by $(-1)$
are also possible but they correspond to some multiple (modulo 5) of the choice
above. Other choices for the block representing $\hat D_4$ within $\tilde D_4$ would
lead to results differing from (\ref{Z5charges}) or its multiple by a $U(1)$
transformation.

We first conclude that the $\Z_5$ charges of $Q_{iL}$ are all different but, in
contrast to the $\Z_3$-2HDM, they do not exhaust all the charges, and it is
not guaranteed anymore that any column of any $\Gamma$ or $\Delta$ has a non-zero
entry.
Nevertheless, all $\Z_5$ charges can be recovered by taking combinations of
$\bar{Q}_{iL}$ with $\phi_a$ or $\tphi_a$.
We can explicitly list the charges for
\eq{
(\bar{Q}_{iL}\phi_1)^*\sim\mtrx{2\cr1\cr0},~~
(\bar{Q}_{iL}\phi_2)^*\sim\mtrx{0\cr4\cr3},~~
(\bar{Q}_{i}\phi_3)^*\sim\mtrx{3\cr2\cr1},
}
and for
\eq{
(\bar{Q}_{iL}\tphi_1)^*\sim\mtrx{2\cr1\cr0},~~
(\bar{Q}_{iL}\tphi_2)^*\sim\mtrx{4\cr3\cr2},~~
(\bar{Q}_{i}\tphi_3)^*\sim\mtrx{1\cr0\cr4}.
}
Extra complex conjugation everywhere is placed to make the comparison with right
quark fields more direct. We see that whatever the $\Z_5$ charge of a right quark
field is, it is always possible to couple it to some $\bar{Q}_{iL}\phi_a$ for $d_R$
or $\bar{Q}_{iL}\tphi_a$ for $u_R$.

Using these expressions, we then define two $5\times 3$ matrices,
\eq{
A_d=\mtrx{(\yukd[2])_2\cr(\yukd[2])_3\cr\hline
	  (\yukd[1])_1\cr(\yukd[1])_2\cr(\yukd[1])_3}
    \sim\mtrx{4\cr3\cr2\cr1\cr0},\qquad
A_u=\mtrx{(\yuku[2])_1\cr(\yuku[2])_2\cr\hline
	  (\yuku[1])_1\cr(\yuku[1])_2\cr(\yuku[1])_3}
    \sim\mtrx{4\cr3\cr2\cr1\cr0},~~
}
where $(\yukd[i])_j$ denotes the $j$-th row of $\yukd[i]$ and the numbers
denotes the corresponding charges of $(\bar{Q}_{iL}\phi_a)^*$ and
$(\bar{Q}_{iL}\tphi_a)^*$ for $A_d,A_u$, respectively.
Each $(\yukd[i])_j$ can have non-zero elements only at positions
which correspond to the field $d_R$ whose $\Z_5$-charge is equal to the value
indicated here.

We arrive at the following recipe to construct all possible textures for $\Z_5$
invariant 3HDM:
\begin{enumerate}
\item Choose a texture for $A_u$ and $A_d$ where each column has
\textit{exactly} one nonzero entry. This choice determines the charges of $d_{iR}$
and $u_{iR}$, and it also completely defines $\yukd[1]$ and $\yuku[1]$
and two rows in $\yukd[2]$ and $\yuku[2]$.
\item The rest of entries of $\Gamma$'s and $\Delta$'s can be extracted by matching
the charges for the right-handed fields.
\item Apply permutations on the rows of $\yukd[i],\yuku[i]$ (permutations
on $Q_{Li}$) and permute $\yukd[i],\yuku[i]$ themselves (permutations on
$\phi_{i}$).
\end{enumerate}
For step (2) we use the correspondence
\eq{
\label{Z5:correspondence}
\mtrx{-\cr\hline(\yukd[3])_1\cr(\yukd[3])_2\cr(\yukd[3])_3\cr\hline(\yukd[2])_1}
    \sim\mtrx{4\cr3\cr2\cr1\cr0},~~
\mtrx{(\yuku[3])_3\cr\hline-\cr\hline
	  (\yuku[2])_3\cr\hline(\yuku[3])_1\cr(\yuku[3])_2}
    \sim\mtrx{4\cr3\cr2\cr1\cr0}.
}
Notice that we can have between 4 and 6 Yukawa terms in each sector depending on whether
the charges of $d_{iR}$ and $u_{iR}$ appear once more in \eqref{Z5:correspondence}.
For example, we have $4$ terms in $d$-sector if $d_{iR}\sim(4,4,2)$.
The possibility of 3 terms is excluded because it leads automatically to massless
quarks.

Let us exemplify the recipe.
We choose 
\eq{
\label{Z5:texture:example}
A_d=\mtrx{0&0&0\cr*&0&0\cr\hline0&0&*\cr0&*&0\cr0&0&0},~~
A_u=\mtrx{0&*&0\cr0&0&0\cr\hline*&0&0\cr0&0&0\cr0&0&*}.
}
Then the charges for $d_{iR}$ and $u_{iR}$ are
\eq{
(d_{iR})\sim\mtrx{3\cr1\cr 2},~~
(u_{iR})\sim\mtrx{2\cr4\cr 0}.
}
The textures for $\yukd[1],\yuku[1]$ can be directly read off from
\eqref{Z5:texture:example}, below the horizontal line, while the rest of the Yukawa
matrices will be combinations in accordance to \eqref{Z5:correspondence}:
\begin{gather}
\yukd[1]=\mtrx{0&0&*\cr0&*&0\cr0&0&0},~~
\yukd[2]=\mtrx{0&0&0\cr0&0&0\cr*&0&0},~~
\yukd[3]=\mtrx{*&0&0\cr0&0&*\cr0&*&0},\cr
\yuku[1]=\mtrx{*&0&0\cr0&0&0\cr0&0&*},~~
\yuku[2]=\mtrx{0&*&0\cr0&0&0\cr*&0&0},~~
\yuku[3]=\mtrx{0&0&0\cr0&0&*\cr0&*&0}.
\end{gather}
This recipe allows one to construct a plenty of examples of 3HDM Yukawa quark
sectors compatible with $\Z_5$ flavor symmetry, the maximal discrete abelian
symmetry for 3HDM. We do not go into a detailed counting of {\em all distinct}
allowed textures.

\section{Maximal symmetries for $N=4$ and $N=5$}

Continuing the investigation of small $N$ cases, 
we now apply the strategy of Sec.\,\ref{sec:3hdm:Z5} to extract the maximal
symmetries for theories with $N=4$, $|G_F|=8$, and $N=5$ Higgs doublets,
$|G_F|=12$.
Some comments on nonmaximal groups which are realizable can be found in appendix
\ref{ap:realizable}.

Let us begin with $N=4$ where the reduced matrix $\tilde D_5$ has $n=5$ rows
$\tbd_i$ of length 7.
In this case, the backbone structure for maximal symmetry will not be unique as it
was for $N=2$ and $N=3$.
As we have seen in Sec.\,\ref{subsection-bound:2}, maximal symmetry is achieved
when the 5 $\bq$-vectors are divided into $(n_1,n_2,n_3)=(2,2,1)$ vectors of each
type in \eqref{set:q+}, as in table \eqref{TableGF:n}.
We can safely suppose $\bq_4=\bq_1$ and $\bq_5=\bq_2$ while $\bq_3$ obeys
\eqref{q3=q1+q2}.
The specific choice for $\bq_1,\bq_2$, could be \eqref{choice:q12}.

By expanding $E(\tbd_1,\tbd_2,\tbd_3,\tbd_4,\tbd_5)$ we obtain the eight terms 
(as before, $k$ stands for $\bp_k$, and $\slashed{k}$ stands for $\bq_k$)
\eqali{
\label{n=5:exp}
E(\tbd_1,\tbd_2,\tbd_3,\tbd_4,\tbd_5)
&=
(\slashed{1}\slashed{2}345)-(\slashed{1}\slashed{2}245)-(\slashed{1}\slashed{2}234)
-(\slashed{1}\slashed{2}145)
\cr&
~~~+\ (\slashed{1}\slashed{2}135)-(\slashed{1}\slashed{2}125)
-(\slashed{1}\slashed{2}124)+(\slashed{1}\slashed{2}123)
\,.
}
To get the maximal symmetry, we need all of them to be nonzero and equal.
After analyzing the possibilities for the $\bp$-vectors, as in
Sec.\,\ref{sec:3hdm:Z5}, we conclude that 
\eq{
\label{n=5:p45}
\bp_4=-\bp_1-\bp_3\,,\quad
\bp_5=-\bp_2-\bp_3,
}
provided that these combinations give allowed $\bp$-vectors. 
The expression (\ref{n=5:exp}) thus simplifies to $8(\slashed{1}\slashed{2}123)$,
and the maximal symmetry is attained by choosing linearly independent
$\bp_1,\bp_2,\bp_3$.
The detailed steps leading to \eqref{n=5:p45} can be found in Appendix
\ref{ap:tech}.

We can make explicit choices for the $\bp$-vectors by rearranging the order of
$\phi_i$ appropriately, thus making use of the transformations of
Sec.\,\ref{subsection-backbone}.
For example, we can fix
\eq{
\label{n=5:p3}
\bp_3=(1,-1,0,0)=\bbe_1-\bbe_2\,,
}
where $(\bbe_i)_j=\delta_{ij}$ are the canonical basis vectors in $\RR^4$.
Now let us focus on the pair of vectors $\bp_1,\bp_4$.
They are related by \eqref{n=5:p45} and restricted to be $\bp$-vectors.
From the fact that $\bp_3+\bp_1$ and $\bp_3+\bp_4$ should be distinct $\bp$-vectors,
we conclude that
\eq{
\label{n=5:set}
\bp_1,\bp_4\in
\{\bbe_2-\bbe_j,\bbe_j-\bbe_1\}\,,
}
where $j\neq1,2$. Since $\bp_1$ and $\bp_4$ are interchangeable because
$\bq_4=\bq_1$, the specific choice is irrelevant and we can choose any order for
\eqref{n=5:set} for a given $j$.
The same reasoning applies to the pair $\bp_2,\bp_5$, except that the index $j=3$ or $4$ has
to be different from the pair $\bp_1,\bp_4$.
Using the freedom to relabel $\phi_i$, we can arbitrarily choose $j=3$ for the pair
$\bp_1,\bp_4$ and $j=4$ for the pair $\bp_2,\bp_5$.
Therefore, we are left with \textit{only one backbone structure for 4HDM with
maximal abelian symmetry}.

In order to find which abelian group of order 8 we get,
we reconstruct the backbone structure and find its Smith normal form. 
If we choose the following ordering for \eqref{n=5:set}, we get
\eq{
\tD_5 = 
\left(
\begin{array}{cccc|ccc}
 0 & 1 & -1 & 0 & 1 & 0 & -1\\
 0 & 1 & 0 & -1 & -1 & 1 & 0\\
 1 & -1 & 0 & 0 & 0 & 1 & -1\\
 -1 & 0 & 1 & 0 & 1 & 0 & -1\\
 -1 & 0 & 0 & 1 & -1 & 1 & 0
\end{array}
\right)\,.
}
The SNF of $\hat{D}_5$, for which we take as usual the central square block, 
is then diag$(1,1,1,1,8)$, which corresponds to $\Z_8$.
The $\Z_8$ charges are calculated using (\ref{RDCphases}) 
and are equal to
\eq{
\ts=(0, 2, 1, -3; -1, 2, 0)^\tp\,.
}

For $N=5$, the same strategy applied for $N=4$ would work without change.
However, the number of terms in the expansion of $\det\hat{D}_6$ and the number of
possibilities increase considerably.
Thus we use a different strategy to narrow down the possibilities and then
survey the reduced possibilities on the computer.

The number of rows in $\hat{D}_6$ is $n=6$. 
The maximal symmetry is attained when we have two $\bq$-vectors of each type.
We can choose $\bq_4=\bq_1$, $\bq_5=\bq_2$, as in the $N=4$ case, and consider
additionally $\bq_6=\bq_3=\bq_1+\bq_2$.
The Higgs part of the rows, $\bp_i$, now has 5 components but we can still fix
$\bp_1$ as 
\eq{
\bp_1=(1,-1,0,0,0)\,.
}
We leave the remaining 5 $\bp$-vectors as generic.
We can now apply row operations on $\tD_6$ to eliminate $\bq_3,\bq_4,\bq_5,\bq_6$.
Only the following nontrivial part is then relevant for the symmetry:
\eq{
\tD_6\to \tD_{4\times 5}=
\mtrx{\bp_3-\bp_1-\bp_2\cr\bp_4-\bp_1\cr\bp_5-\bp_2\cr\bp_6-\bp_3}\,.
}
Note that neither of the rows can vanish.

We now survey all possible $\tD_{4\times 5}$ by varying the vectors $\bp_i$,
$i=2,3,4,5$. This procedure leads to 48 matrices corresponding to the maximal
symmetry $\ZZ_2\times\ZZ_6$, which has order 12.
All of these matrices can be shown to be equivalent up to transformations from
Sec.\,\ref{subsection-backbone}. For example, interchanging $\bp_2$ and $\bp_5$
is irrelevant because $\bq_2=\bq_5$.
Thus, we are left with \textit{only one backbone structure for 5HDMs with the
maximal abelian symmetry.}
One possible form for $\tD_6$ is
\eq{
\tD_6
=
\left(
\begin{array}{ccccc|ccc}
 1 & -1 & 0 & 0 & 0 & 1 & 0 & -1\\
 1 & 0 & -1 & 0 & 0 & -1 & 1 & 0\\
 0 & 0 & 0 & 1 & -1 & 0 & 1 & -1\\
 0 & 1 & 0 & -1 & 0 & 1 & 0 & -1\\
 0 & 0 & 1 & -1 & 0 & -1 & 1 & 0\\
 -1 & 0 & 0 & 0 & 1 & 0 & 1 & -1
\end{array}
\right)\,.
}
The SNF in this case is diag$(1,1,1,1,2,6)$, and the $\Z_2\times\Z_6$ generators
are
\eq{
\ts_{\Z_2}=(0, 0, 1, 0, 1; 0, 1, 0)^\tp,~~
\ts_{\Z_6}=(0,1, 1, 2, 4; 1, 2,0)^\tp\,.
}
Note that in this case the abelian group corresponding to the maximal symmetry 
is not cyclic.

We finally arrive at the following two conclusions:
\begin{itemize}
\item
all the upper bounds on the second row of \eqref{TableGF} are achievable 
and this upper bound is exact at least up to $N=5$;
\item
in each of these cases, the NHDM with the maximal symmetry 
has quark Yukawa matrices which are determined by {\em only one backbone structure}.
\end{itemize}
The explicit construction of these matrices can be done using the procedure explained with the cases
$N=2$ and $N=3$.

\section{Conclusions}

Explaining the flavor puzzle in terms of hidden symmetries in the flavor sector
which become broken by scalar fields is an intriguing and actively pursued idea.
Nature provides us with the results, --- the fermionic mass and mixing patterns, ---
but does not give any hint of the possible flavor symmetry group which may be at work.
It is not surprising that in this situation many works simply pick up a specific group 
or a series of groups and explore their phenomenological consequences.

With this paper, we bring some order to this activity, at least, if the scalar
sector consists of several electroweak doublets. 
We do not advocate specific groups for solution of the flavor puzzle, 
but we give a recipe where to stop in search for such groups and
how to analyze consequences of any of its abelian subgroups.

Specifically, we develop a set of tools which
allows one, for any chosen number of Higgs doublets $N$, to write down the short
list of {\em all} allowed finite abelian groups,
and for each group, to explicitly build {\em all} quark Yukawa textures compatible with it
and not immediately leading to gross contradictions with experiment
(namely, no massless quarks and no massless scalars).
It generalizes the results (but not the methods!) of the work \cite{FS2011} 
from 2HDM to general $N$, 
and it is competitive and, arguably, simpler than the recent
alternative work \cite{Serodio:2013gka}.
In particular, we find a simple upper bound, Eq.\,\eqref{TableGF}, on the
order of the realizable finite abelian groups, show that this upper bound is achievable for $N\le
5$, and prove that these maximal abelian symmetries 
are connected to only one basic structure in each case.
Testing whether the models which we construct are actually quantitatively compatible with observed
numerical values of the quark masses, the entries of the CKM matrix, and the current
limits on New Physics requires a dedicated study, which is clearly beyond the scope
of this work.

On a different note, we suggest to view this work as a particular application
of the very general and powerful method of using Smith normal forms for
the determination of the rephasing symmetry of a model.
Although we considered only Yukawa interaction of quarks with Higgs doublets,
the method itself is completely universal and can be applied to any collection
of interacting complex fields and any form of interaction Lagrangian.

This method and its potential power are poorly known in the HEP community,
which is rather surprising because the task of identifying an abelian symmetry group
of a model is very natural.
In fact, many works proposed methods which seem to bear some resemblance
to the Smith normal form approach, without developing its full potential.
We hope that this approach can find other applications within HEP and, especially,
in building models of New Physics.

\section*{Acknowledgements} 
The work of I.P.I.\ was supported in part by the grant RFBR 11-02-00242-a,
the RF President Grant for scientific schools NSc-3802.2012.2,
and the Program of the Department of Physics
of the Scientific Council of the Russian Academy of Sciences
``Studies of Higgs boson and exotic particles at LHC.''
I.P.I. also acknowledges the support of FAPESP through grant 2012/23040-5
and 
is thankful for hospitality to UFABC, where this work started.

The work of C.C.N. was partially supported by Brazilian CNPq and Fapesp.

\appendix
\section{Realizable symmetries}
\label{ap:realizable}

In the main text we focused on variants of the NHDM with 
maximal abelian symmetries.
In this Appendix we comment on nonmaximal groups and show that most of them 
are also realizable, at least for $N$ up to 5.

First, we show that any discrete symmetry $G$ which is realizable for $N$
Higgs doublets is also realizable for $N+1$ doublets.
This can be done by explicitly constructing the reduced matrix $\tD_{n+1}$,
from $\tD_n$, possessing the same (discrete) symmetry as $\tD_n$.
We construct $\tD_{n+1}$ by adding one column of zeros in the first column
of $\tD_n$ and one row in the $(n+1)$-th position of the form
\eq{
\label{d:n+1}
\tilde{d}^{(n+1)}=(\bp_{n+1};\bq_{n+1})=(\pm1,\cdots).
}
Here dots indicate that the remaining entries can be arbitrary; they just need to comply with the
general principles from which the vectors  $\bp_{n+1}$ and $\bq_{n+1}$ are built.
The process of transforming the upper-right $\tD_n$ block of $\tD_{n+1}$ into its
SNF does not change the first entry of $\tilde{d}^{(n+1)}$ in \eqref{d:n+1}.
In that form, we can rearrange columns to make $\tD_{n+1}$ lower triangular. 
The SNF of $\tD_{n+1}$ will have the same integers as $\tD_{n}$ with one additional
unity.
This construction works for any $N$. In particular, even for $N>5$, where the rows
of $\tD_n$ can not be generic, we can take the first entry of \eqref{d:n+1} to be
$\pm1$ from the contribution of $A$ in \eqref{D:ABC}, which is generic.

Therefore one strategy for complete classification of realizable discrete
abelian groups would be to survey the possible symmetries for $N+1$ which
are not possible for $N$.

By using such a strategy, we can already fill some gaps.
By taking $\bp_4=-\bp_1$ in \eqref{D4:exp:2}, instead of \eqref{n=3:p4}, we can
conclude that $|G_F|=4$ is realizable for $N=3$ ($n=4$), and the SNF confirms that
the group is $\Z_4$.
The group $\Z_2\times\Z_2$ of same order can be obtained by using $\bq_3=\bq_2$,
instead of \eqref{q3=q1+q2}, and by considering $\bp_4=-\bp_1$ together with
$\bp_3=-\bp_2$.
This implies that all abelian groups up to order 5 are realizable for the
Yukawa sector of the 3HDM.

For $N=4$, we can get the symmetry $\Z_7$ by replacing the second relation of
\eqref{n=5:p45} with $\bp_5=\bp_1-\bp_2$. 
By replacing the second relation of \eqref{n=5:p45} with $\bp_5=-\bp_2$ we obtain
$\Z_6$.
One can also obtain $\Z_6$ by using $(n_1n_2n_3)=(320)$ instead of the maximal
option $(221)$ in \eqref{TableGF:n}. 
Therefore, all cyclic groups with order up to $8$ are realizable for the
Yukawa sector of 4HDM. 
We find, however, that the noncyclic groups, $\Z_2\times\Z_4$ and
$(\Z_2)^3$, are not realizable. The unique realizable noncyclic group is
$\Z_2\times\Z_2$.

For $N=5$, we could partially survey the realizable abelian groups. We already
exclude from the groups of maximal order the possibility of
$\Z_{12}\simeq\Z_4\times\Z_3$.
Seeking for groups of nonmaximal order, which do not appear for $N=4$, we have
explicitly checked the possibility for $\Z_2{\times}\Z_4$, $\Z_3\times \Z_3$, $\Z_9$, 
$\Z_{10}$, and $\Z_{11}$.

\section{Technical calculation}
\label{ap:tech}

We show here the detailed steps to deduce the relations in Eq.\,\eqref{n=5:p45}.
We need to analyze the possibilities for the $\bp$-vectors considering that
they can be only of the form $(1,-1,0,0)$ up to permutations; hence they belong to
a subspace of dimension 3.
The last term of \eqref{n=5:exp} is only nonzero if $\bp_1,\bp_2,\bp_3$ are
linearly independent. Thus we can use them to decompose $\bp_4,\bp_5$ as
\eq{
\bp_4=a_1\bp_1+a_2\bp_2+a_3\bp_3\,,\quad
\bp_5=b_1\bp_1+b_2\bp_2+b_3\bp_3\,.
}
Non cancellation among the last three terms of \eqref{n=5:exp} implies $a_3=b_3=-1$.
To keep the sign of the third term equal to the last term, we need $a_1=-1$.
The same reasoning for the fifth term implies $b_2=-1$.
From the fact that the second and fourth terms should be equal to the last term
(due to Lemma of Sec.\,\ref{subsection-bound}), we deduce $a_2=b_1=0$.
We finally arrive at \eqref{n=5:p45}.


\end{document}